\documentclass[9pt,twocolumn,twoside]{optica}
\setboolean{shortarticle}{false}
\setboolean{minireview}{false}

\title{ML-SIM: A deep neural network for reconstruction \\of structured illumination microscopy images}

\author[1,2]{Charles N. Christensen}
\author[1]{Edward N. Ward}
\author[2]{Pietro Lio}
\author[1,*]{Clemens F. Kaminski}
 
\affil[1]{University of Cambridge, Department of Chemical Engineering and Biotechnology, Laser Analytics Group, Cambridge, UK}
\affil[2]{University of Cambridge, Department of Computer Science and Technology, Artificial Intelligence Group, Cambridge, UK}

\affil[*]{Corresponding author: cfk23@cam.ac.uk}
\affil[{}]{Interactive demo and source code: \url{http://ML-SIM.com}}

\usepackage{booktabs}
\usepackage{multirow}

\begin{abstract}
Structured illumination microscopy (SIM) has become an important technique for optical super-resolution imaging because it allows a doubling of image resolution at speeds compatible for live-cell imaging. However, the reconstruction of SIM images is often slow and prone to artefacts. Here we propose a versatile reconstruction method, ML-SIM, which makes use of machine learning. The model is an end-to-end deep residual neural network that is trained on a simulated data set to be free of common SIM artefacts. ML-SIM is thus robust to noise and irregularities in the illumination patterns of the raw SIM input frames. The reconstruction method is widely applicable and does not require the acquisition of experimental training data.  Since the training data are generated from simulations of the SIM process on images from generic  libraries the method can be efficiently adapted to specific experimental SIM implementations.  The reconstruction quality enabled by our method is compared with traditional SIM reconstruction methods, and we demonstrate advantages in terms of noise, reconstruction fidelity and contrast for both simulated and experimental inputs. In addition, reconstruction of one SIM frame typically only takes ~100ms to perform on PCs with modern Nvidia graphics cards, making the technique compatible with real-time imaging. The full implementation and the trained networks are available at \url{http://ML-SIM.com}.
\end{abstract}

\begin{document}

\maketitle

\section{Introduction}
\label{sec:int}

Structured illumination microscopy (SIM) is an optical super-resolution imaging technique that was proposed more than a decade ago \cite{sheppard1988super,Heintzmann1999,Gustafsson2000,3D-SIM,3D-SIM2}, and continues to stand as a powerful alternative to techniques such as Single Molecule Localization Microscopy (SMLM) \cite{Moerner1989,Betzig1995} and Stimulated Emission Depletion (STED) microscopy \cite{Hell1994}. The principle of SIM is that by illuminating a fluorescent sample with a patterned illumination, interference patterns are generated that contain information about the fine details of the sample structure that are unobservable in diffraction-limited imaging. In the simplest case of a sinusoidal illumination pattern with a spatial frequency of $k_0$, the images acquired are a superposition of three copies of the sample's frequency spectrum, shifted by +$k_0$, 0, and -$k_0$. The super-resolution image is reconstructed by isolating the three superimposed spectra and shifting them into their correct location in frequency space. The resulting spectrum is then transformed back into real space, leading to an image that is doubled in resolution. Isolating the three frequency spectra is mathematically analogous to solving three simultaneous equations. This  requires the acquisition of three raw images, with the phase of the SIM patterns shifted with respect to one another along the direction of $k_0$. Ideally these phase shifts are in increments of $2 \pi /3$ to ensure that the averaged illumination, i.e. the sum of all patterns, yields a homogeneous illumination field. Finally, to obtain isotropic resolution enhancement in all directions, this process is repeated twice, rotating the patterns by $2 \pi /3$ each time, to yield a total of 9 images (i.e. 3 phase shifts for each of the 3 pattern orientations).
\begin{figure*}[t!]
  \centering
  \includegraphics[width=1\textwidth]{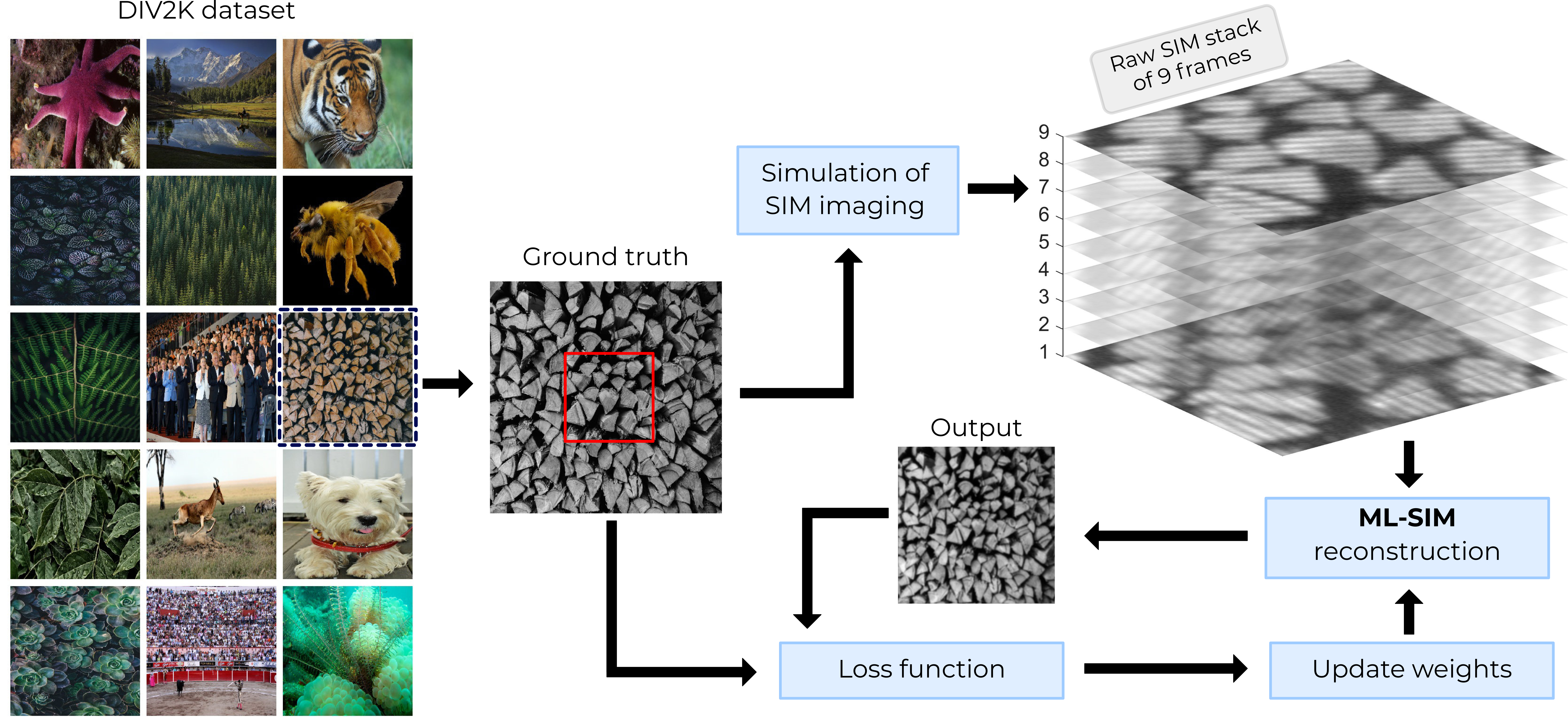}
  \caption{Data processing pipeline for ML-SIM. Training data for the model is generated by taking images from the commonly used image dataset DIV2K and simulating the imaging process of SIM using a model adapted from the open source library OpenSIM. The simulation can be further optimised to reflect the properties of the experimental system for which the reconstruction method is desired, for example to match the pixel size of the detector or numerical aperture of the detection optics. The outputs of the simulation are image stacks of the same size as those acquired by the microscope (here 9 frames).}
  \label{fig-pipeline}
\end{figure*}
While SIM can be extended to resolve features down to the 50-60 nm range \cite{li2015extended,rego2012nonlinear}, it does not offer the highest resolution of the available super-resolution methods. However, the inherent speed of SIM makes it uniquely suited to live-cell imaging \cite{Strohl2016b,winter2014faster}. SIM also requires relatively low illumination intensities, and therefore reduced photo-toxicity and photo-bleaching compared to other methods. Many of the drawbacks of SIM relate to the reconstruction process, which can be time consuming and prone to artifacts. In all but optimal imaging conditions, deviations from the expected imaging model or the incorrect estimation of experimental  parameters (pixel size, wavelength, optical transfer function, image filters, phase step size etc.) introduce artefacts, degrading the final image quality \cite{SIM-check}. This becomes especially prominent for images with low signal-to-noise ratios, where traditional methods will mistakenly reconstruct noise as signal leading to artefacts that can be hard to distinguish from real features in the sample. At worst, the  reconstruction process  fails completely. These issues can introduce an element of subjectivity into the reconstruction process, leading to a temptation to adjust reconstruction parameters until the 'expected' result is obtained. In addition, traditional reconstruction methods are computationally demanding. The processing time for a single reconstruction in popular  implementations such as FairSIM \cite{Muller2016}, a plugin for ImageJ/Fiji, and OpenSIM running in MATLAB \cite{Lal2016}, can reach tens of seconds even on high-end machines, making real-time processing during SIM image acquisition infeasible. Finally, traditional methods cannot easily reconstruct images from SIM data that is underdetermined, e.g. inputs with fewer than 9 frames and / or recordings with uneven phase steps between frames . These drawbacks limit the applicability of SIM when imaging highly dynamic processes \cite{Strohl2017b}. Examples include the peristaltic movement of the endoplasmic reticulum \cite{holcman2018single} or the process of cell division \cite{planchon2011rapid}, which require low light level imaging at high speed to reduce the effects of photo-toxicity and photo-bleaching. 

In this work we propose a versatile reconstruction method, ML-SIM, that addresses these issues. The method uses an end-to-end deep residual neural network that is trained to reduce  reconstruction artefacts affecting traditional methods.  The method is robust against noise and irregularities in the illumination patterns used for producing raw SIM data frames. This is possible because in ML-SIM, training data are generated synthetically by a simulation of the SIM imaging process. Simulating the data provides ideal targets (ground truths) for supervised learning, which lets the model train without being exposed to traditional reconstruction artefacts. Although the training data used here are simulated and unrelated to real microscopic samples, we find that the model generalises well and we demonstrate successful application to experimental data obtained from two distinct SIM microscopes. This greatly empowers the method in the context of super-resolution SIM imaging, since obtaining ground truth data is difficult, if not impossible, to obtain.  Furthermore, since the training data are simulated in ML-SIM, the reconstruction model is easily retrained, if needed. This means that the model can be customised to SIM setups of any configuration by changing simulation parameters used in the  generation of the training data. 

\begin{figure*}[t!]
  \centering
  \includegraphics[width=1\textwidth]{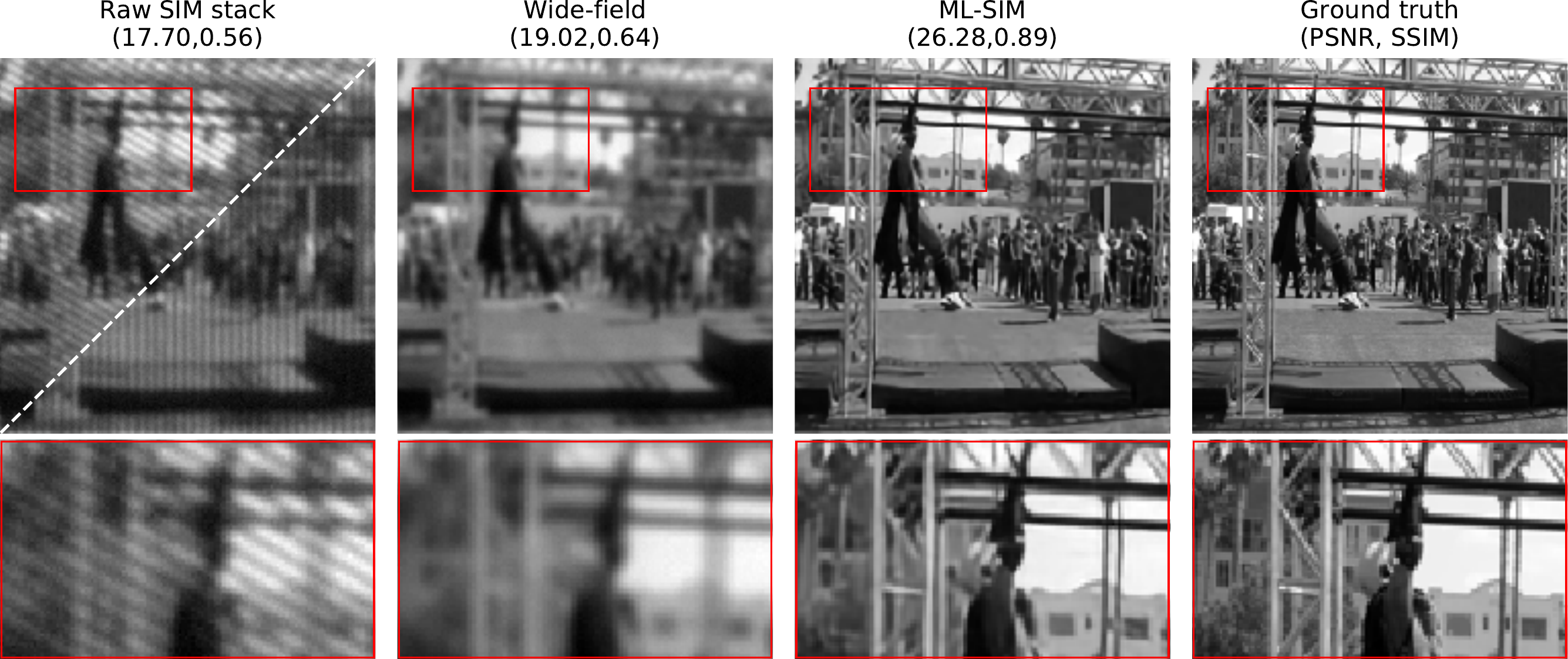}
  \caption{Generation of training data sets for ML-SIM. 1st column: Sample from test partition of DIV2K (ground truth) transformed to a raw data stack of 9 frames via simulation of the SIM imaging process.  Two different orientations are shown for the excitation patterns. Column 2: Wide-field image, obtained as the mean of the 9 raw frames. Column 3: Super-resolved image obtained on reconstruction with ML-SIM. Column 4: Ground truth. The image quality metrics shown in brackets are the peak signal-to-noise ratio and the structural similarity index \cite{wang2004image}, respectively. }
  \label{fig-div2ktestresult}
\end{figure*}

\section{Results}
\label{sec-res}
\subsection{Using machine learning to train a reconstruction model}
ML-SIM is built on an artificial neural network. Its purpose is to process a sequence of raw SIM frames (i.e. a stack of nine images representing the number of raw images acquired in a SIM experiment), into a single super-resolved image. To achieve this, a supervised learning approach is used to train the network, where pairs of inputs and desired outputs (ground truths) are presented during training to learn a mapping function. These training data could be acquired on a real SIM system using a diverse collection of samples and experimental conditions. However, the targets corresponding to the required inputs are more difficult to obtain. At least two approaches seem possible: (A) using outputs from traditional reconstruction methods as targets; and (B) using images from other super-resolution microscopy techniques that are able to achieve higher resolution than SIM (e.g SMLM or STED). Option (A) would prohibit ML-SIM from producing reconstructions that surpass the quality of traditional methods and would be prone to reproduce the artifacts mentioned in Section \ref{sec:int}. Option (B) requires a capability to perform correlative imaging of the same sample, which may be difficult to achieve, since training requires 
hundreds or even thousands of distinct data pairs \cite{doi:10.1021/acs.chemrev.6b00604}. In addition, both  approaches require the preparation of many unique samples to build a  training set diverse enough for  the model to generalise well. Hence,  these options were not pursued in this work and we approached the problem instead by starting with ground truth images and  simulating inputs by mimicking the SIM process in silico, allowing for very diverse training sets to be built. We used the image set DIV2K \cite{Agustsson2017}, which consists of 1000 high-resolution images of a large variety of objects, patterns and environments. To generate the SIM data, images from the image set were first resized to a standard resolution of 512x512 pixels and transformed to greyscale. Raw SIM images were then calculated using a SIM model adapted from  the OpenSIM package  \cite{Lal2016}. The model and underlying  parameters are described in Section \ref{sec-met}. The simulated raw SIM stacks were used as input to the neural network and the output compared to the known ground truth in order to calculate a loss to update the network weights. Figure \ref{fig-pipeline} shows an overview of the training process with an example of a simulated SIM input. The architecture of the neural network is further described in Section \ref{sec-met}.

\begin{figure*}[t!] 
  \centering
  \includegraphics[width=1\textwidth]{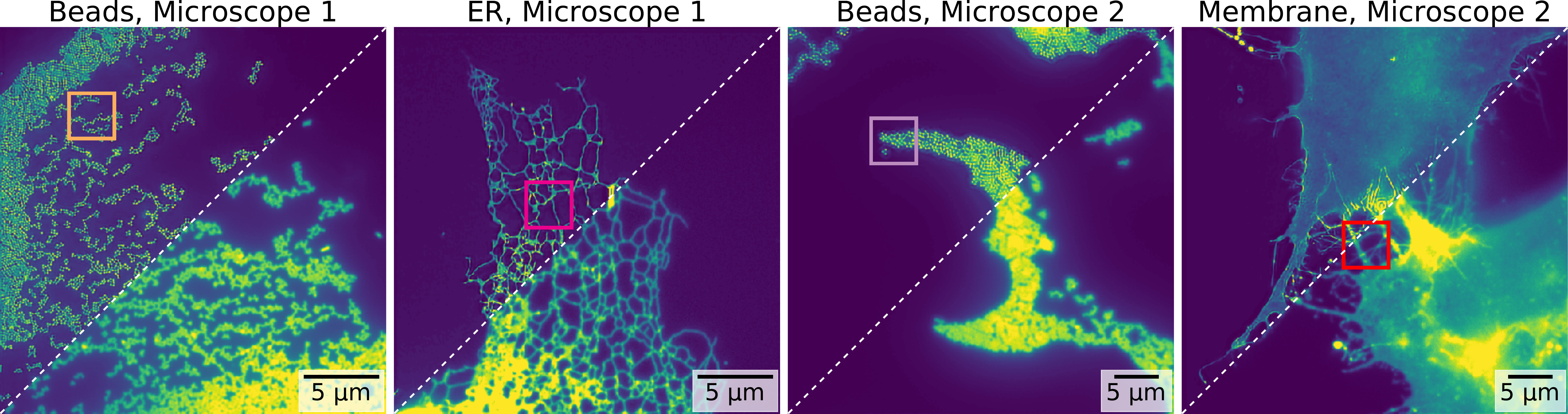} 
  \\ \vspace{5pt}
  \includegraphics[width=1\textwidth]{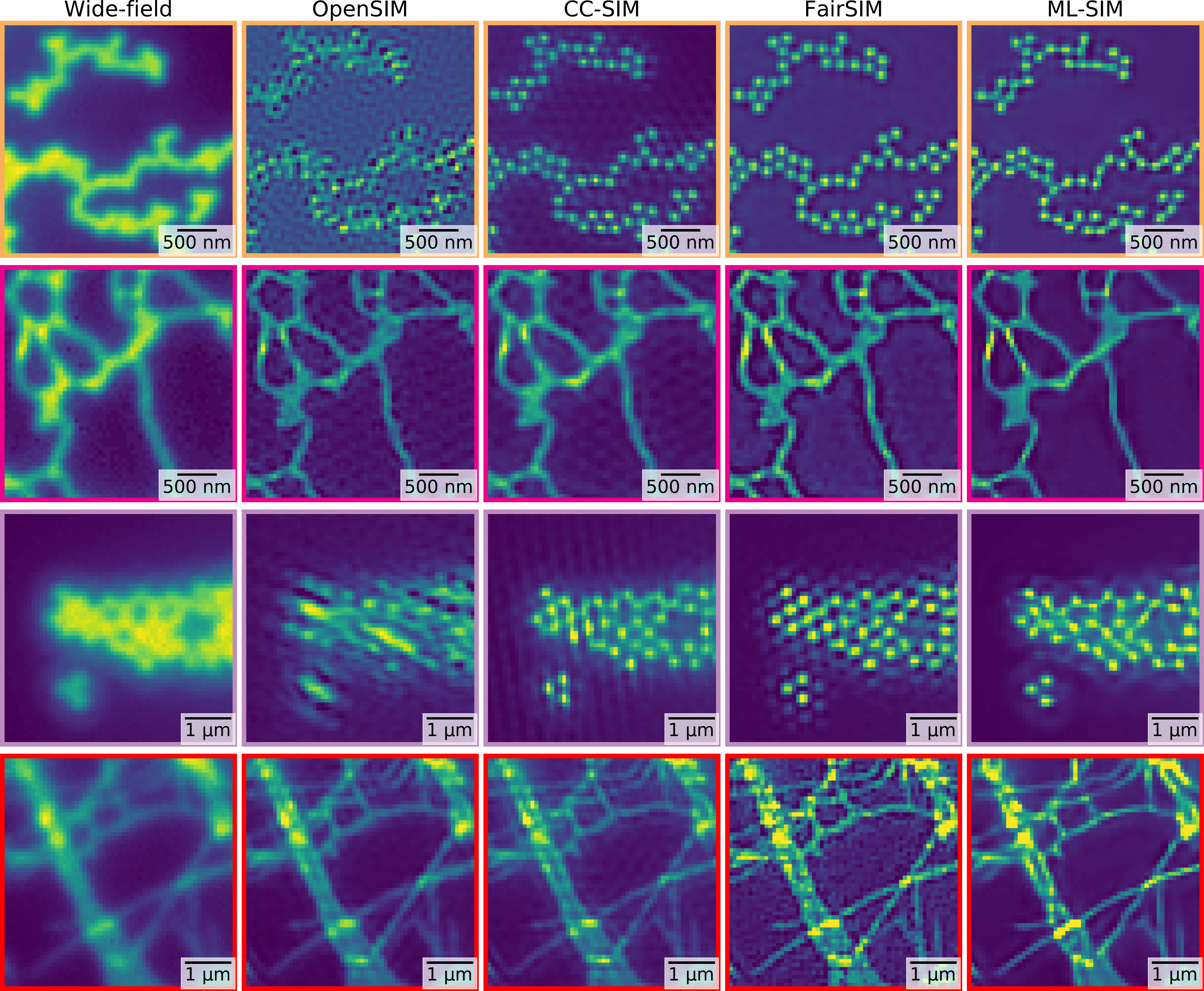} 
  \caption{Reconstruction of SIM images from four different samples imaged on two different experimental SIM set-ups. Microscope 1 uses a spatial light modulator for stripe pattern generation \cite{fiolka2008structured}, while microscope 2 uses interferometric pattern generation. Both instruments were used to image a sample consisting of fluorescent beads as well as biological samples featuring the endoplasmic reticulum and a cell membrane, respectively. (Top) Full field-of-view images where each upper left half shows the reconstruction output from ML-SIM and each lower right half shows the wide-field version taken as the mean of the raw SIM stack.
  (Bottom) Cropped regions of reconstruction outputs from OpenSIM \cite{Lal2016}, CC-SIM \cite{Wicker2013}, FairSIM \cite{Muller2016} and ML-SIM. Panels in rows 2 to 5 correspond to regions indicated by coloured boxes in the full frame images.}
  \label{fig-slm}
\end{figure*}

  \begin{figure*}[t!] 
    \centering
    \includegraphics[width=1\textwidth]{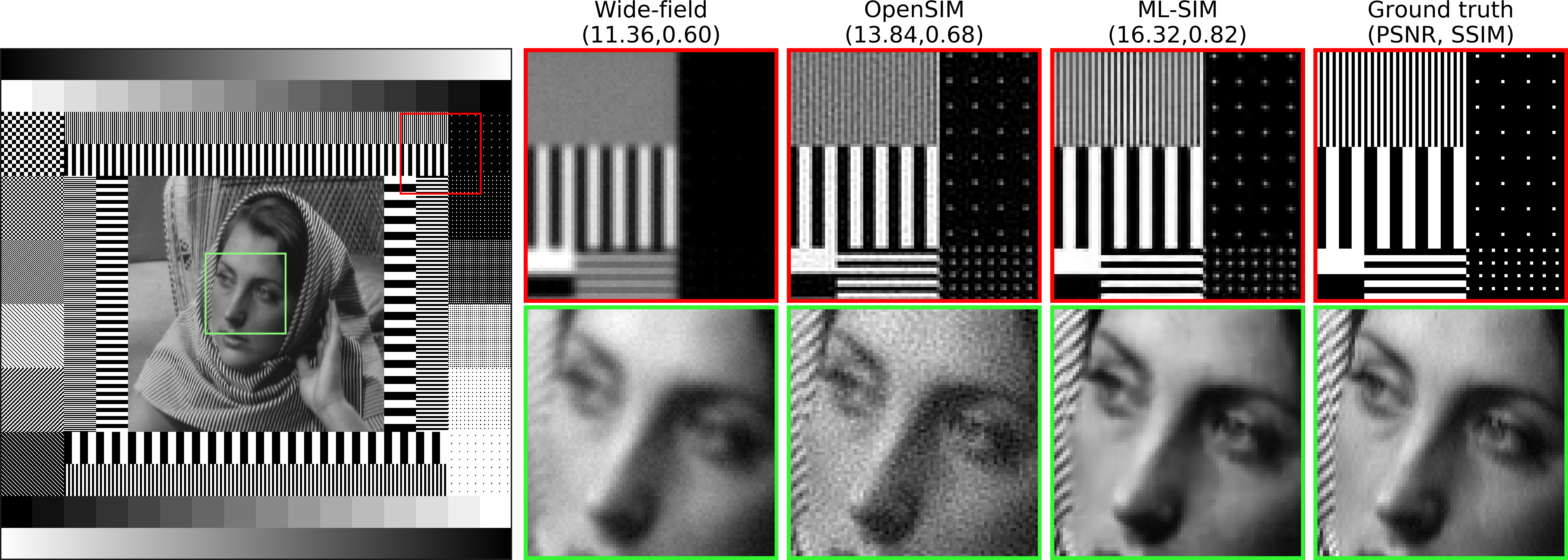} 
    \caption{Reconstructions of a test target with OpenSIM and ML-SIM and comparison to the ground truth. OpenSIM was found to be the best performing traditional method on this test sample, both in terms of PSNR and SSIM with the other methods achieving PSNR scores of 12.56 dB (CC-SIM) and 12.88 dB (FairSIM).  }
    \label{fig:barbara}
  \end{figure*}

\subsection{Application of the trained model}
To begin with, we tested that the network had learned to reconstruct simulated SIM stacks. Prior to training, a separate partition of DIV2K was selected for testing. A sample from this test partition  is shown on Figure \ref{fig-div2ktestresult}. The stripe pattern for two of the nine frames of the input SIM stack are shown in the leftmost panel. The stripe patterns cancel out when all 9 frames are summed together (second column), and this corresponds to the case of even illumination in a wide-field microscope. Compared to the wide-field image, the reconstruction from ML-SIM is seen to have a much improved resolution with a peak signal-to-noise ratio (PSNR) value more than 7 dB higher as well as a significantly higher structural similarity index (SSIM). Beyond these metrics, several features of the image can  be seen to be resolved after reconstruction that were not visible before hand, such as the vertical posts seen to the right side of the cropped region. 

It should also be noted that the reconstruction has not amplified the noise by introducing any evident artefacts, even though the input image featured a significant amount of Gaussian noise in addition to randomisation of the stripe frequency and phase -- see Section \ref{sec-met} for definitions of those parameters. As further described in Section \ref{sec-met}, the neural network underlying ML-SIM is different to those of generative networks, which means that the model is more strongly penalised during training for introducing image content that is not in the real image. We argue that, even though this results in slightly more blurred output images than would be achievable with a generative network\cite{Ledig2016,Wang2018}, the absence of artificial features is preferable in scientific imaging applications. This trade-off is referred to as minimising the mathematical reconstruction error (e.g. root-mean-square deviation) rather than optimising the perceptual quality \cite{Blau2018,Vasu2018}. 

While ML-SIM is able to reconstruct simulated SIM stack inputs, it is of course only valuable if it also works on real SIM data, acquired experimentally. The ML-SIM model was trained on input data from simulations, using data bearing little resemblance to real-world biological SIM data. Any success for real world SIM reconstructions therefore requires the  model to have generalised the SIM process in such a way that it it becomes independent of image content and sample type. This requires a realistic simulation of the SIM imaging process to generate training data, that is sufficiently diverse on the one hand, and  reflects measurement  imperfections as encountered in practical SIM imaging. The former was avoided through use of a diverse training dataset, and the latter through use of the well-known imaging response function (Section \ref{sec-met}, Equation \eqref{eq-irf}), and introduction of uncertainty in the stripe patterns. To test ML-SIM on experimental data, SIM images of different samples were acquired with two different SIM setups \cite{JOVE-SIM}. The resulting reconstructed outputs are shown on Figure \ref{fig-slm}, where they are compared to outputs of traditional reconstruction methods: OpenSIM \cite{Lal2016}, a cross-correlation (CC-SIM) phase retrieval approach \cite{Wicker2013, matlabCC-SIM}, and FairSIM \cite{Muller2016}. ML-SIM is seen to obtain resolution on par with the other methods but producing less noisy background and fewer artefacts. The bottom two rows of images of beads and cell membranes were acquired with phase steps deviating from the ideal $2 \pi / 3$.  This reflects a difficulty with the interferometric SIM setup (see Section \ref{sec-met}) to achieve equidistant, and precisely defined, phase steps for each illumination pattern angle. This means that the reconstruction algorithm must handle inconsistent phase changes, a factor only the cross-correlation method was capable of handling. However, although CC-SIM has improved resolution,  artefacts are apparent, seen as vertical lines and ringing in the images. ML-SIM on the other hand reconstructed with fewer artefacts and strongly improved background rejection.

\subsection{Performance assessment}
We performed a quantitative comparison of ML-SIM with traditional reconstruction methods on reconstructions of simulated raw SIM stacks generated from two image data sets; a subset of 10 unseen DIV2K images and 24 images from a dataset referred to as Kodak 24, commonly used for image restoration benchmarking \cite{Agustsson2017,Lehtinen2018}. Parameters for OpenSIM, CC-SIM and FairSIM were all systematically adjusted to produce the highest achievable output quality. Consequently, each method required completely different parameter configurations than those used for reconstructions of the experimental data shown in Figure \ref{fig-slm}.  For ML-SIM however there were no tunable parameters. The optical transfer function (OTF) is estimated within each method even though the function is known for the simulated images -- this is the same premise as for the reconstruction of the experimental samples on Figure \ref{fig-slm}, for which the OTFs were unknown. Each method applies an identical Wiener filter to the final reconstruction output, whereas the output of ML-SIM is untouched. The performance scores for all methods measured in PSNR and SSIM averaged over the entire image sets are listed in Table \ref{tab:per} with scores for wide-field as a reference. In terms of both metrics, OpenSIM and ML-SIM have the highest scores with a PSNR ~2 dB higher for ML-SIM. CC-SIM and FairSIM lag behind but both methods still succeed in improving the input beyond the baseline wide-field reference. The performance gap between OpenSIM and the other traditional methods is likely due to a better estimation of the OTF, because OpenSIM assumes an OTF that is similar to the one used when simulating the SIM data. 
 \begin{figure*}[t!]
\begin{minipage}[t]{0.365\textwidth}
\includegraphics[width=\textwidth]{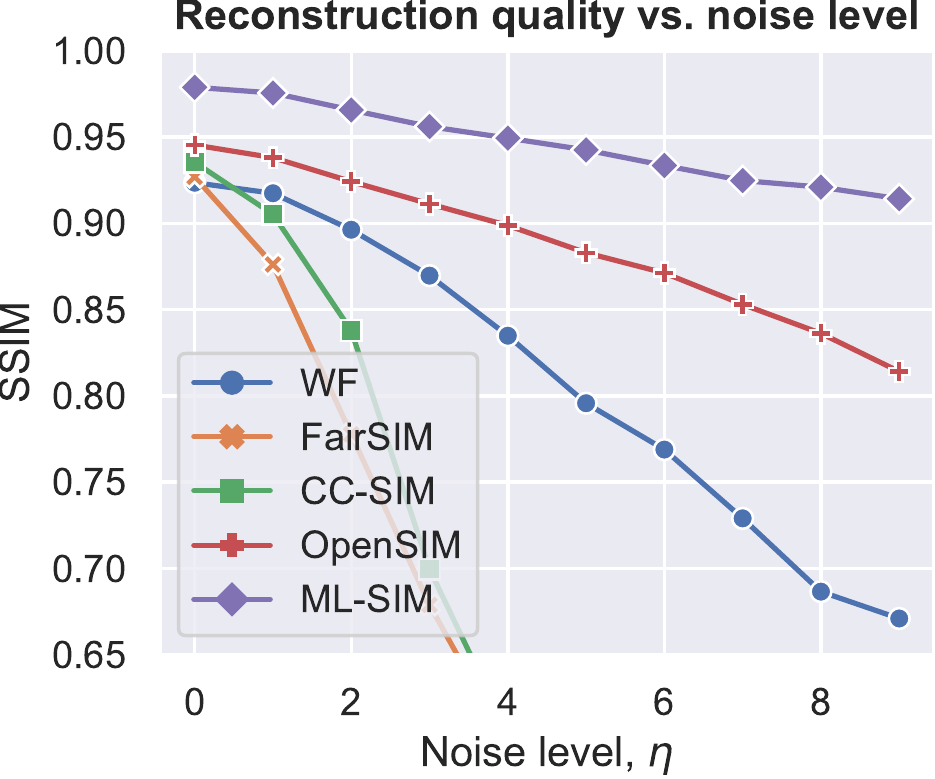}
\end{minipage}
\hspace{10pt}
\begin{minipage}[t]{0.605\textwidth}
\includegraphics[width=\textwidth]{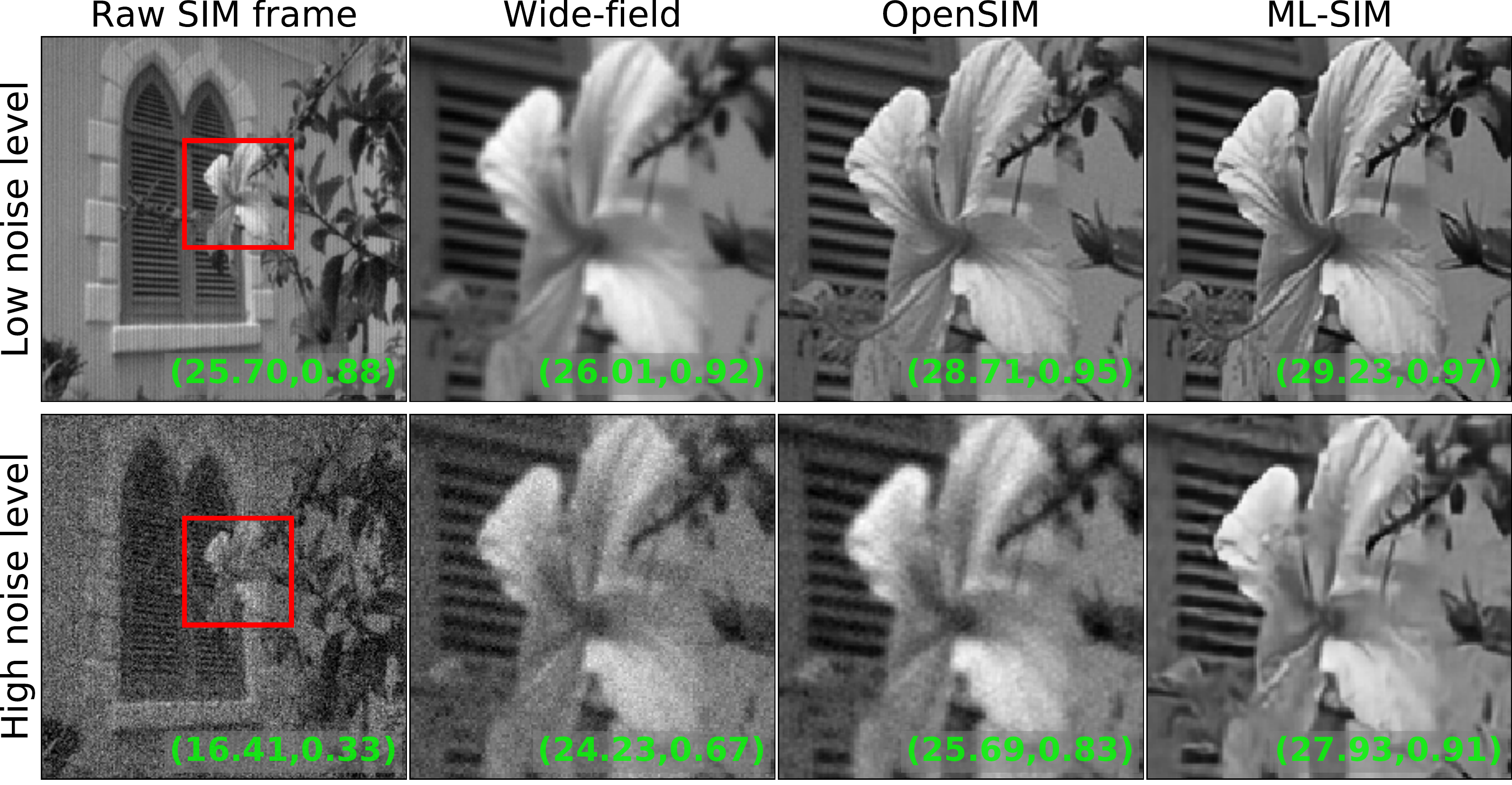}
\end{minipage}
\caption{(Left) Reconstruction quality as measured by the structural similarity index, SSIM, as a function of the amount of noise added to the input images. Gaussian noise is added to every frame of the raw SIM stack. Noise is normally distributed with a standard deviation $\eta\cdot \sigma$, where $\sigma$ is the standard deviation of the input image. (Right) Images at low ($\eta = 0$) and high noise levels ($\eta = 9$) reconstructed with OpenSIM and ML-SIM, respectively. PSNR and SSIM scores using the ground truth as reference are shown in the lower right hand corner of every image.      }
\label{fig-noi}
\end{figure*}

\begin{table}[b!]
  \centering
  \begin{tabular}{l|cc|cc}
          & \multicolumn{2}{c}{\underline{\textbf{DIV2K Test Set}}} & \multicolumn{2}{c}{\underline{\textbf{Kodak 24 Set}}} \\
          & PSNR [dB]        & SSIM        & PSNR [dB]         & SSIM         \\
          \hline
          Wide-field & 25.31       & 0.84       & 24.05         & 0.85        \\
          CC-SIM  & 25.37       & 0.89       & 24.61         & 0.86        \\
          FairSIM & 25.34       & 0.86       & 25.32         & 0.86       \\
          OpenSIM & 28.46       & 0.91       & 27.36         & 0.92        \\
          ML-SIM  & \textbf{30.30}       & \textbf{0.95}       & \textbf{30.22}         & \textbf{0.96}       
  \end{tabular}
  \caption{Test scores on simulated raw SIM data generated from  image sets DIV2K and Kodak 24 for commonly used reconstruction methods and for ML-SIM.}
  \label{tab:per}
  \end{table}
A more challenging test image than those based on DIV2K and Kodak 24 images is shown on Figure \ref{fig:barbara}. This simulated test image is reconstructed with the same three traditional methods. OpenSIM is found to achieve the best reconstruction quality of the three with a PSNR score of 13.84 dB versus 12.56 dB for CC-SIM and 12.88 dB for FairSIM. The same image reconstructed with ML-SIM results in a PSNR score of 16.32 dB -- again about 2 dB higher than OpenSIM. Two cropped regions comparing OpenSIM and ML-SIM are shown on Figure \ref{fig:barbara}. The area in the upper right corner of the test image is particularly challenging to recover due to the single pixel point patterns and the densely spaced vertical lines. While the points vanish in the wide-field image, these are recovered both by OpenSIM and  ML-SIM. The resolution of the point sources are slightly superior in the ML-SIM reconstruction, and ML-SIM manages to recover the high-frequency information in the top line pattern very well. Overall it is also seen that the reconstruction from ML-SIM contains much less noise, which is especially evident in the zoomed region of the face. This suggests that ML-SIM is less prone to amplify noise present in the input image. We tested this further by gradually adding more Gaussian image noise to the input image, and again comparing the reconstructions from the various methods. The results of this test are shown on Figure \ref{fig-noi}, where it is seen clearly that ML-SIM performs best at high noise levels. As more noise is added the gap in performance is seen to increase between ML-SIM and the other models indicating that the neural network has learned to perform denoising as part of the reconstruction process. This is supported by the cropped regions on the right side of the figure, which show cleaner detail in the image when compared to the input, wide-field and OpenSIM images. OpenSIM was found to perform consistently well in this noise test, whereas FairSIM and CC-SIM struggled to reconstruct at all for higher noise levels. This is not surprising,  since added noise may cause the parameter estimation to converge to incorrect optima, which can heavily corrupt the reconstruction outputs. As a result, the reconstructions from FairSIM and CC-SIM were of worse quality than the wide-field reference at higher noise levels.

\section{Discussion}
We demonstrate and validate a SIM reconstruction method, ML-SIM, which is built on a residual neural network. The method is robust to noise and eliminates the need for user-defined parameters. The method was trained by simulating raw SIM image data from images obtained from common image repositories, serving as ground truths.  ML-SIM was found to generalise well and successfully reconstructed  artificial test targets that were of a completely different nature than the natural images used to generate the training datasets. More importantly, it successfully reconstructed real data obtained by two distinct experimental SIM implementations. We compared the performance of ML-SIM to widely used reconstruction methods, OpenSIM \cite{Lal2016}, FairSIM \cite{Muller2016}, and CC-SIM \cite{Wicker2013}. In all cases, reconstruction outputs from ML-SIM contained less noise and fewer artefacts, while  achieving similar resolution improvements. Through a randomsation of phase shifts in the simulated training data, it was also possible to successfully reconstruct images, that could not be processed successfully with two of the traditional reconstruction methods. ML-SIM shows a robustness to unpredictable variations in the SIM imaging parameters and deviations from equidistant phase shifts. Similarly, ML-SIM reconstructed images that were strongly degraded by noise even beyond  the point where the other methods failed. 

An important part of the ML-SIM method is that the underlying neural network is trained on simulated data. This makes the method applicable to any SIM setup regardless of system-specific configurations and without  the need for building paired datasets (i.e. acquiring corresponding raw inputs and ground truths), since the simulation to generate training datasets easily can be adapted to reflect specific setups. 

A future direction could be to fine-tune the training data by incorporating a more sophisticated image formation model. Currently, out-of-focus light from above and below the focal plane is not simulated in the training data. As with the other reconstruction methods, this can result in artefacts in regions of the sample with dense out-of-focus structures. Given that the spatial frequency information required to remove this background is available in SIM, it is possible that an updated ML-SIM network could be constructed that incorporates an efficient means for background rejection \cite{OS-SIM, OS-SIMNeil}.

\begin{figure*}[t!] 
  \centering
  \includegraphics[width=1\textwidth]{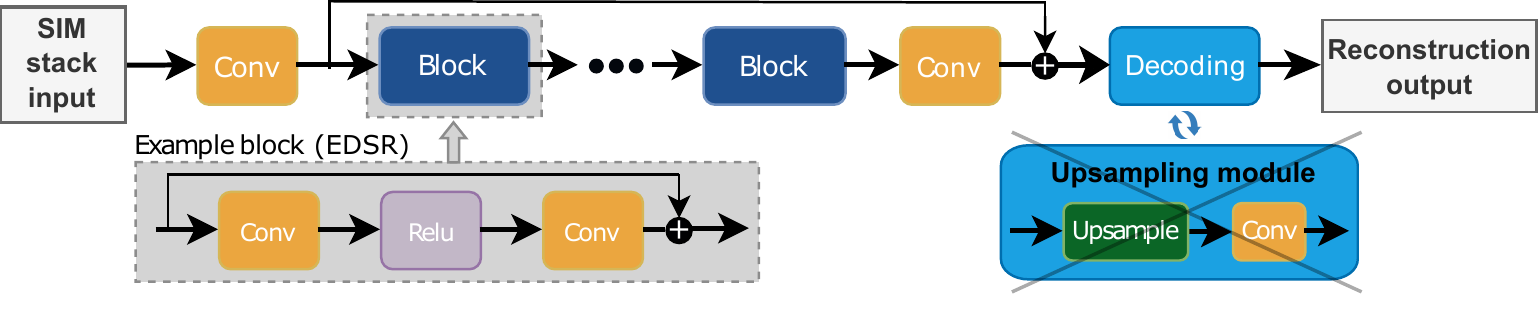} 
  \caption{The architecture of ML-SIM is inspired by state-of-the-art single image super-resolution architectures. This figure shows the architecture of EDSR \cite{Lim2017}, but the same structure applies to RCAN only with a more complex block called a channel attention block. ML-SIM has a RCAN architecture without an upsampling module and with a larger input layer that handles 9 frames. }
  \label{fig:eds}
\end{figure*}

\section{Methods}
\label{sec-met}

\subsection{Online tool and source code}
\label{sec-src}
The source code for training ML-SIM and applying the model for reconstruction is available in a public repository on GitHub -- \href{https://github.com/charlesnchr/ML-SIM}{https://github.com/charlesnchr/ML-SIM}. This repository includes source code for generating the training data by simulating the SIM imaging process with parameters that can be easily adapted to reflect specific SIM setups  (e.g. by changing stripe orientations, number of frames, etc.). We have created a website \href{http://ML-SIM.com}{http://ML-SIM.com} with a browser-based online implementation of ML-SIM that is ready for testing by users with no prior SIM experience.  

\subsection{Convolution neural networks}
\label{sec-cnn}
Artificial neural networks consist of a sequence of layers that each perform a simple operation, typically a weighted sum followed by an nonlinear activation function, where every weight corresponds to a neuron in the layer. The weights are trainable, meaning that they are updated after every evaluation of an input performed during training. The updating scheme can be as simple as gradient descent with gradients determined via backpropagation of a loss calculated as the deviation between the network's output and a known target. A convolutional layer is no different, but utilises spatial information by only applying filters to patches of neighbouring pixels. The number of learned filters in one layer is a parameter, but is typically a power of 2, such as 32, 64 or 128. The network links past layers to present layers by skip connections to avoid the vanishing gradient problem. This type of architecture is known as a residual neural network \cite{He2015}. 

\begin{figure}[h!] 
  \centering
  \includegraphics[width=1\linewidth]{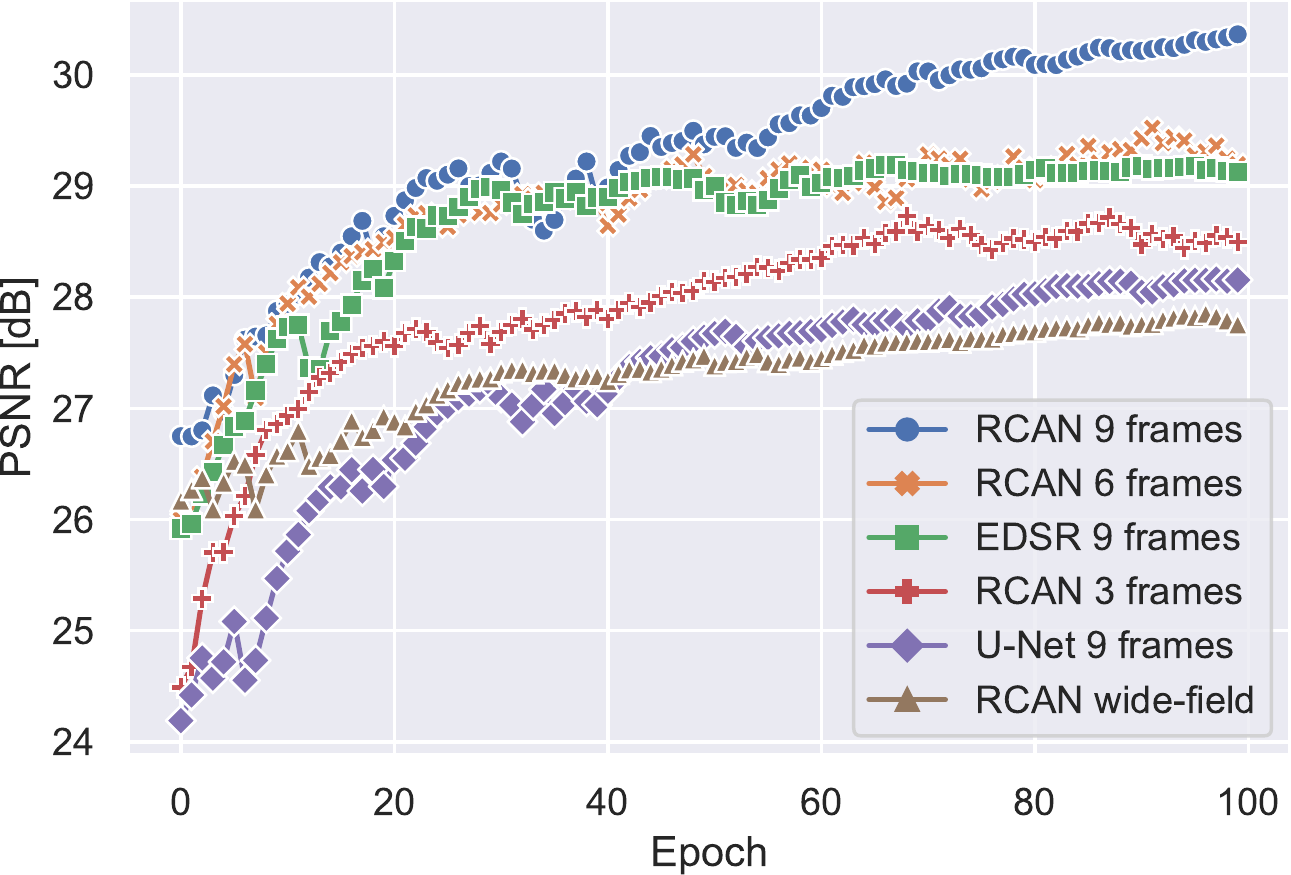} 
  \caption{Validation test set scores during training for different network architectures and input dimensions. The two state-of-the-art single image super-resolution architectures, RCAN and EDSR, have been modified to perform SIM reconstruction. The number of frames of the raw SIM stack, up to a total of 9, is also varied to confirm that the network learns to to extract information from all images in the stack versus a subset of it or the mean of its frames, i.e. inputting wide-field images (RCAN wide-field).     }
  \label{fig-per}
\end{figure}

Several architectures were tested as part of this research to  select the one most suitable for ML-SIM. U-Net \cite{Ronneberger2015} is a popular, versatile and easily trained network, but its performance was found to fall short of state-of-the-art single image super-resolution networks such as EDSR \cite{Lim2017} and RCAN \cite{Zhang2018a}. Those super-resolution networks have been customised to be able to handle input stacks of up to 9 frames and output a single frame with no upsampling, i.e. the upsampling modules of those networks have been omitted -- see Figure \ref{fig:eds} for a depiction. In addition to testing different network architectures the number of frames of the input raw SIM stack, up to a total of 9, was also varied. In Figure \ref{fig-per} the convergence of test scores on a validation set during training are shown for the various architectures and input formats considered. It is found that SIM reconstruction with  subsets containing only 3 or 6 frames still performed significantly better than if the network learns to perform a more simple deconvolution operation by just training on a wide-field input. Only using a subset of 3 frames does however cause a substantial reconstruction quality loss compared to using 6 frames, which is not surprising since the corresponding analytical reconstruction problem becomes underdetermined for fewer than 4 frames \cite{Strohl2017b}. The RCAN model performs better than EDSR with a consistently higher PSNR score when trained on all 9 frames, while performing similarly to EDSR when trained with 3 fewer frames.

Motivated by the results summarised in Figure \ref{fig-per} and with the certainty that the entire input stack is utilised for the output reconstruction, the RCAN architecture was chosen for ML-SIM. The depth of the network was chosen to be around 100 convolutional layers (10 residual groups with 3 residual blocks). The network was then trained for 200 epochs with a learning rate of $10^{-4}$, halved after every 20 epochs, using the Adam optimiser \cite{Kingma}. The models were implemented with Pytorch and trained using a Nvidia Tesla K80 GPU for approximately a day per model. Models have been trained on the full DIV2K imageset except for a small collection of randomly selected images used for validation during training. 

\begin{figure*}[t!]
  \centering
  \includegraphics[width=1\textwidth]{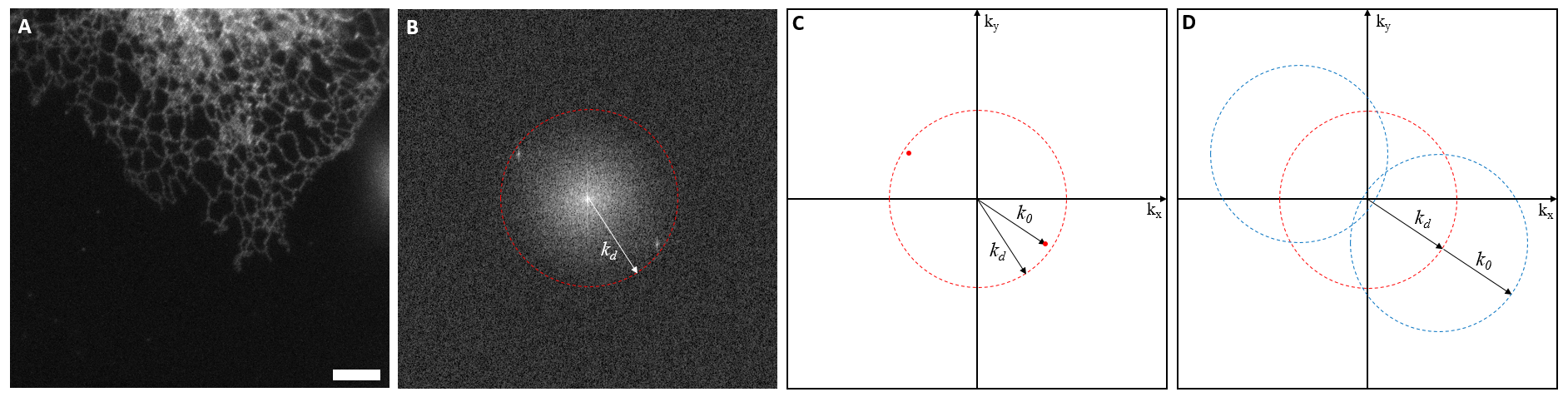}
  \caption{SIM methodology visualised in frequency space. A) Raw image captured during SIM. Scale bar is 5$\mu$m B) 2D Fourier transform of A. The resolution limit can be visualised as a cutoff frequency $k_d$ beyond which no spatial frequency information from the sample is collected. The frequency components of the striped illumination pattern are visible as bright peaks close to the cutoff frequency. C) The frequency components of the excitation pattern, $k_0$, are chosen to be as close to the diffraction limit as possible, to maximise resolution increase. The interference of the patterned illumination with the sample pattern means the observed region of frequency space now contains frequency components from outside the supported region, shifted by $\pm k_0$. D) By shifting the phase of the pattern, the regions of frequency space can be isolated and moved to the correct location in frequency space. The maximum spatial frequency recovered is now $k_d$ + $k_0$.}
  \label{fig-simmethod}
\end{figure*}
\subsection{Generating simulated data}

The ideal optical transfer function is generated based on a given objective numerical aperture (NA), pixel size and fluorescence emission wavelength. The illumination stripe patterns were then calculated from their spatial frequency $k_0$ and a phase $\phi$,
\begin{align}
  \label{eq-pat}
  I_{\theta,\phi}(x,y) = I_0 \left[ 1 - \frac{m}{2}\cos \left( 2\pi( k_x \cdot x + k_y \cdot y) + \phi \right) \right],
\end{align}
where $k_x$, $k_y$ = $k_0 \cos \theta$, $k_0 \sin \theta$ for a pattern orientation $\theta$ relative the the horizontal axis. $\phi$ defines the phase of the pattern (i.e. the lateral shift in the direction of $k_0$) and $m$ is the modulation factor which defines the relative strength of the super-resolution information contained in the raw images. In total, 9 images were generated for each target image, corresponding to three phase shifts for each of three pattern orientations -- see Figure \ref{fig-simmethod} for a depiction. The fluorescent response of the sample can then be modelled by the multiplication of the sample structure, $S(x,y)$ (input image), and the illumination pattern intensity $I_{\theta,\phi}(x,y)$. The final image, $D_{\theta,\phi}(x,y)$, is then blurred by the PSF, $H(x,y)$, and noised with the addition of white Gaussian noise, $N(x,y)$,
\begin{align}
  \label{eq-irf}
  D_{\theta,\phi}(x,y) = \left[ S(x,y) I_{\theta,\phi} \right] \otimes H(x,y) + N(x,y),
\end{align}
where $\otimes$ is the convolution operation. In addition to the Gaussian noise, $N(x,y)$, added pixel-by-pixel, a random error is added to the parameters for the stripe patterns, $k_0$, $\theta$ and $\phi$, to approximate the inherent uncertainty in an experimental setup for illumination pattern generation. 

\subsection{Microscopy}

For the experimental data described in Section \ref{sec-res}, two custom-built SIM microscopes were used. For the imaging of the endoplasmic reticulum a SIM instrument based on a phase only spatial light modulator was used. The microscope objective used was a 60X 1.2 NA water immersion lens and fluorescence was imaged with a sCMOS camera. Cells were labelled with VAPA-GFP and excited by 488 nm laser light.
For the imaging of the cell membrane, a novel SIM setup based on interferometry for the pattern generation was used. In this system the angle and phase shifts are achieved by rotation of a scanning mirror, the repeatability of which introduces uncertainty into the phase shifting. The microscope objective used was a 60X 1.2 NA water immersion lens and fluoresence was imaged with a sCMOS camera. The cell membrane was stained with a caax-venus label and excited with 491 nm laser light. 
On both systems, 200 nm beads labelled with Rhodamine B were excited by 561 nm laser light. 
For both images the traditional reconstruction methods that have been tested -- i.e. OpenSIM, the cross-correlation method and FairSIM -- managed to reconstruct the raw SIM stacks, although with varying success for the interferometric SIM setup due to the irregularity of the phase stepping. 



\section*{Acknowledgments}

The authors thank Jerome Boulanger from the Laboratory of Molecular Biology in Cambridge, UK, for general guidance on reconstruction for SIM. The authors also thank group members Meng Lu and Lisa Hecker from the Laser Analytics Group of the University of Cambridge for providing experimental data from the SLM and interferometric SIM microscopes described in this study. CFK acknowledges funding from the UK Engineering and Physical Sciences Research Council, EPSRC (grants EP/L015889/1 and EP/H018301/1), the Wellcome Trust (grants 3-3249/Z/16/Z and 089703/Z/09/Z) and the UK Medical Research Council, MRC (grants MR/K015850/1 and MR/K02292X/1), MedImmune, and Infinitus (China) Ltd.

\section*{Disclosures}

\medskip

\noindent\textbf{Disclosures.} The authors declare no conflicts of interest.

\bibliography{library}
\bibliographyfullrefs{library}

\end{document}